%% file: main.tex
\documentclass[conference]{IEEEtran}
\IEEEoverridecommandlockouts

\AtBeginDocument{%
  \providecommand\BibTeX{{%
    \normalfont B\kern-0.5em{\scshape i\kern-0.25em b}\kern-0.8em\TeX}}}

\usepackage{amssymb}
\usepackage{xcolor}

\usepackage{import}
\usepackage{algpseudocode,algorithm}
\usepackage{listings}
\usepackage{multirow}
\usepackage{caption}
\usepackage{subcaption}
\usepackage{adjustbox}
\begin{document}

\title{\huge Sequence-Based Incremental Concolic Testing of RTL Models \thanks{This work was partially supported by the Semiconductor Research Corporation (2020-CT-2934)}}

 \author{Hasini Witharana, Aruna Jayasena and Prabhat Mishra \\ 
 University of Florida, Gainesville, Florida, USA}


\maketitle

\begin{abstract}
Concolic testing is a scalable solution for automated generation of directed tests for validation of hardware designs. Unfortunately, concolic testing also fails to cover complex corner cases such as hard-to-activate branches. In this paper, we propose an incremental concolic testing technique to cover hard-to-activate branches in register-transfer level models. We show that a complex branch condition can be viewed as a sequence of easy-to-activate events. We map the branch coverage problem to the coverage of a sequence of events. We propose an efficient algorithm to cover the sequence of events using concolic testing. Specifically, the test generated to activate the current event is used as the starting point to activate the next event in the sequence. Experimental results demonstrate that our approach can be used to generate directed tests to cover complex corner cases while state-of-the-art methods fail to activate them.  


\end{abstract}

\pagestyle{empty}

\input{sections/intro.tex}

\input{sections/relate.tex}

\input{sections/concolic.tex}
\input{sections/select.tex}
\input{sections/experiment.tex}
\input{sections/conclusion.tex}


\bibliographystyle{unsrt}
\bibliography{reference}

\end{document}

%% file: sections/intro.tex
\section{Introduction}

Functional validation is a major bottleneck for modern System-on-Chip (SoC) designs. According to the Wilson Research 2020 functional verification study~\cite{fosterwilson}, more than 50\% of development time in hardware designs were spent in verification. Simulation is the most widely used form of functional validation. Even millions of random tests may not be able to activate complex corner cases such as hard-to-detect branches in Register-Transfer Level (RTL) designs. As a result, it is unlikely to achieve 100\% functional coverage using random or constrained-random tests for industrial designs. To improve the coverage, verification engineers typically write manual tests to cover the remaining functional scenarios. Manual test writing can be cumbersome and error-prone.  There is a critical need for automated generation of directed tests.

\begin{figure}[t]
	\centering
\vspace{-0.1in}
\includegraphics[width=1\columnwidth]{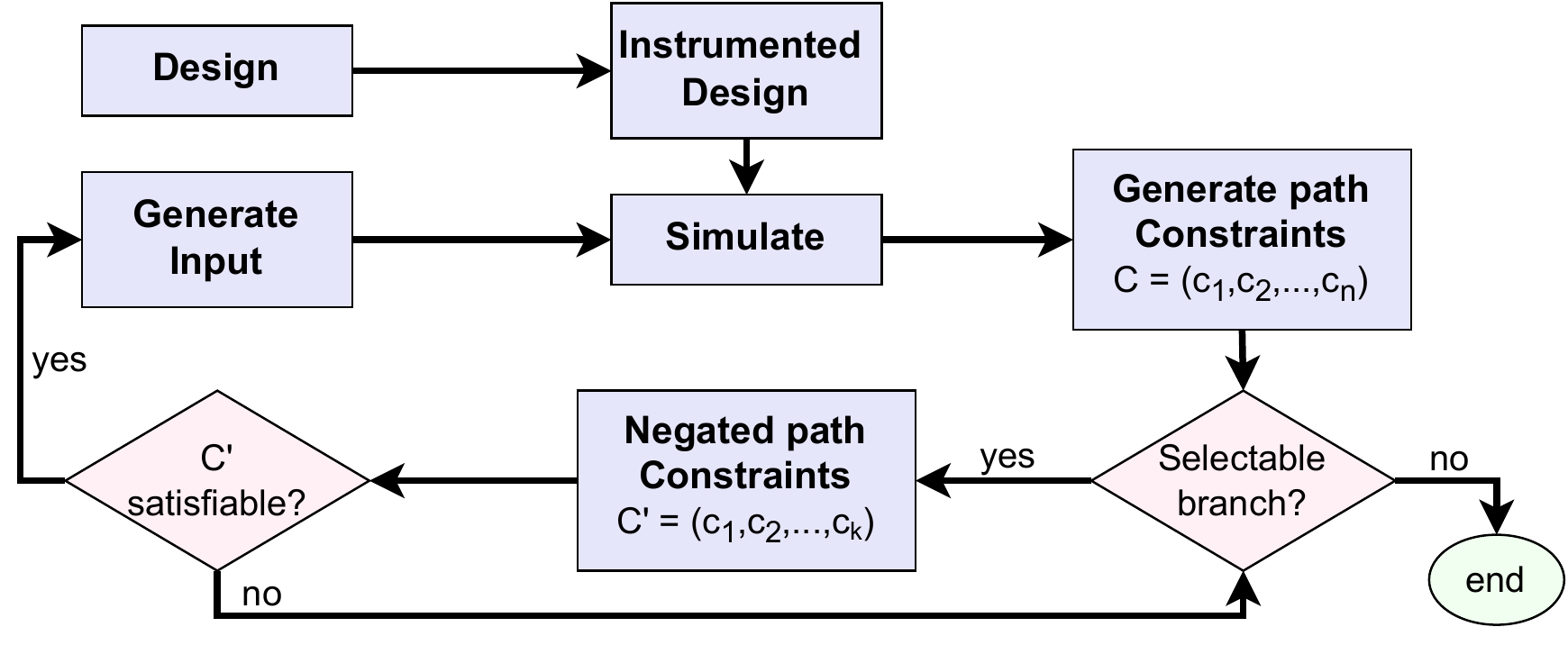}
 \vspace{-0.3in}
	\caption{An overview of concolic testing}
	\label{fig:concolic}
	    \vspace{-0.2in}
\end{figure}

Concolic testing has been successfully used as a directed test generation method in both software~\cite{godefroid2005dart,sen2006cute} and hardware domains~\cite{lyu2020scalable}. Figure~\ref{fig:concolic} shows an overview of the concolic testing framework. The design is instrumented so that the tool can identify  the executed path during simulation. Next, the instrumented design is simulated using an initial vector. The initial test vector can be generated using random or any other test generation methods. The execution path of the design is identified by analyzing the simulation trace. Next, an alternate path is selected by negating one of the branch constraints. The path constraints to activate the selected branch (alternate branch) will be sent to a constraint solver. Constraint solver will produce a solution if the constraints are satisfiable. This solution is used to generate a new test vector to activate the selected branch. If the constraint solver cannot solve the constraints (solution is unsatisfiable), an alternate branch is selected. This process continues until the expected coverage is achieved. Since concolic testing explores one path at a time, it overcomes the state space exploration problem. However, concolic testing faces the path exploration problem due to the exponential number of possible paths to explore. Path exploration problem can be mitigated by using a profitable alternate branch selection approach.

    \begin{figure*}[htp]
	\centering
\vspace{-0.1in}
\includegraphics[width=2.0\columnwidth]{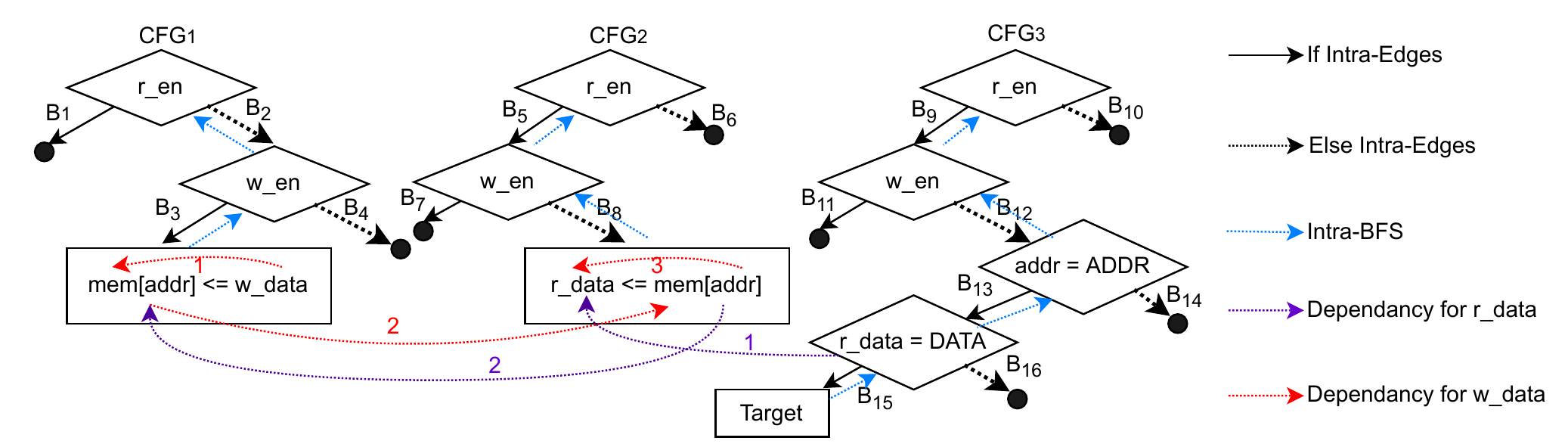}
 \vspace{-0.15in}
	\caption{Control and data flow graphs for the $ram$ design in Listing~\ref{lst:example1}. (BFS: Breadth First Search)}
	\label{fig:cfg}
	    \vspace{-0.15in}
\end{figure*}

Alternate branch selection depends on the coverage goal. Existing approaches try to maximize the overall coverage~\cite{ahmed2017quebs} as well as specific branch targets~\cite{lyu2020scalable, lyu2019automated}. In this paper, we are considering activation of corner cases which are hard-to-activate. Some branches become hard-to-activate due to the complex temporal dependencies that should be preserved in-order to activate that branch. Consider line 36 in Listing 1 that reads a value ($r\_data$) from a specific memory address ($addr)$. For this condition to be true, a write should happen to that specific memory address with the exact values. The read can only happen when read flag ($r\_en$) is true and write flag ($w\_en$) is false. However, write can only proceed when read flag ($r\_en$) is false and write flag ($w\_en$) is true. These are contradictory constraints  that must be satisfied in-order to activate the branch. Existing concolic testing~\cite{lyu2020scalable} fails unless the design is sufficiently unrolled in such cases. Unrolling for a large number of cycles is not feasible for large designs. 

In this paper, we propose a sequence-based incremental concolic testing. Our proposed technique uses edge exploration by traversing the Control Flow Graph (CFG) of the RTL design to identify the event sequence. Next, it solves each sequence while maintaining the order and preserving each solution for solving the next sequence incrementally. This paper makes the following three major contributions.
\begin{enumerate}
    \item Proposes an event sequence based approach for concolic testing. 
    For a given branch, the sequence of events is identified by statically analyzing the concurrent CFGs of the RTL design.
    \item Incrementally applies concolic testing on an event sequence and preserves the test vectors to build the directed test to activate the target (corner case).
    \item Extensive experimental evaluation using a memory design demonstrates the effectiveness of our approach.
\end{enumerate}


%% file: sections/relate.tex
\vspace{-0.1in}
\section{Related Work}
\label{sec:background}



Concolic testing is a promising alternative to model checking based test generation. Specifically, it provides an effective combination of concrete simulation and symbolic execution~\cite{lyu2020scalable}. Unlike model checking that tries to explore
all possible (exponential) execution paths at the same
time, concolic testing explores only one execution path at a time. Concolic testing has been successfully applied on both software~\cite{godefroid2005dart,sen2006cute,majumdar2007hybrid,artzi2010directed} and hardware designs~\cite{liu2009star, liu2014scaling,lyu2020scalable}.
Although concolic testing can avoid state explosion problem, it faces path explosion problem since it needs to select a profitable path is each iteration. While there are promising solutions for selecting beneficial branches~\cite{lyu2020scalable}, they are not suitable for complex corner cases such as hard-to-detect branches with complex branch conditions. We propose an efficient mechanism to activate complex branch conditions by identifying it as a sequence of simple conditions and incrementally applying concolic testing to activate these simple conditions.

%% file: sections/concolic.tex
\section{Incremental Concolic Testing}
\label{sec:concolic}

\begin{figure*}[htp]
	\centering
	\footnotesize
\vspace{-0.1in}
\includegraphics[width=1.8\columnwidth]{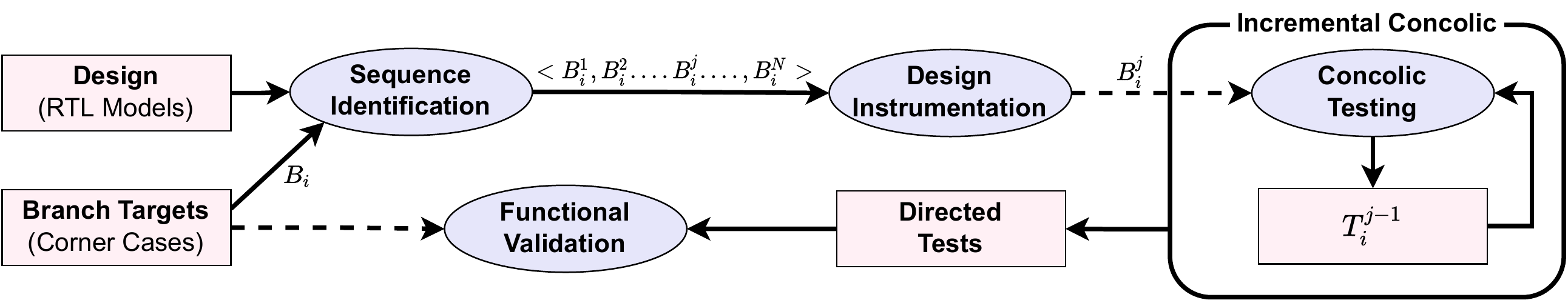}
\vspace{-0.1in}
	\caption{Overview of our framework that consists of sequence identification, instrumentation and incremental concolic testing.}
	\label{fig:overview}
	    \vspace{-0.2in}
\end{figure*}

Figure~\ref{fig:overview} presents an overview of our proposed incremental concolic testing framework. It consists of three major tasks: sequence  identification, design instrumentation, and incremental concolic testing. Algorithm~\ref{alg:overview} shows the relation between the three tasks. Given a design (D) and a branch target ($B_i$), the first step is to identify the set of sequences ($SS$) such that $B_i \rightarrow$  $<S_1, S_2,  ....,  S_n>$. The second step is to instrument the design by converting each sequence to a branch statement. The second step results in instrumented design ($iD$) and the target queue ($TQ$). The third step is to apply concolic testing for each of the branch statements in the order of the sequence. The generated test can be used to activate the branch target during functional validation.

\vspace{0.05in}
\noindent {\bf Example 1:}
We use a simple Verilog design (Listing~\ref{lst:example1}) to describe various concepts in this paper. Listing~\ref{lst:example1} has three $always$ blocks corresponding to three functionalities in a simple memory module: write functionality (line 9 - 18), read functionality (line 19 - 28), and system functionality (line 29 - 42). While read and write are basic memory operations, the system functionality can be viewed as the top module (e.g., processor) trying to check a write followed by a read. For the ease of illustration, we are not showing all the else blocks for the $if$ statements.  $\Box$

\vspace{-0.1in}
\begin{lstlisting}[frame=single, caption={Example of a memory module in Verilog}, label={lst:example1}, language=Verilog, basicstyle=\footnotesize]]
1. module ram
2.  input                   clk, rst,
3.  input  [ADDR_W-1:0]     addr,
    //write signals
4.  input                   w_en,
5.  input [DATA_W-1:0]      w_data,
    //read signals
6.  input                   r_en,
7.  output reg [DATA_W-1:0] r_data
    //memory declaration
8.  reg [DATA_W-1:0]  mem [2**ADDR_W-1:0];

//Memory write
9.  always @(posedge clk) begin
10.  if(r_en) begin
11.   //B1
12.  end
13.  else begin 
14.   if(w_en)begin //B2
15.     mem[addr] <= w_data; //B3
16.   end
17.  end
18. end

//Memory read
19. always @(posedge clk)  begin
20.  if(r_en)
21.    if (w_en) begin //B5
22.     //B7
23.    end
24.    else begin
25.     r_data <= mem[addr]; //B8
26.    end
27.  end
28. end

//Check write followed by read
29. always @(*) begin
30.   if(r_en) begin
31.     if (w_en) begin  //B9
32.      //B11
33.     end
34.     else begin
35.      if(addr == ADDR) begin //B12
36.         if (r_data == DATA) begin //B13
37.           $display("Target"); //B15
38.         end
39.       end
40.     end
41.   end
42. end
43. endmodule
\end{lstlisting}

Figure~\ref{fig:cfg} presents the control and data flow for Listing~\ref{lst:example1}. The three always blocks presented in the example corresponds to the three CFGs as $CFG_1$ (memory write), $CFG_2$ (memory read), and $CFG_3$ (check). The solid black lines represent control flow when the branch condition is true, while the flow for the false condition is represented using black dotted lines. 

\begin{algorithm}[htp]
\footnotesize
\caption{Sequence-Based Incremental Concolic Testing}\label{alg:overview}
\begin{flushleft}

\hspace*{\algorithmicindent} \textbf{Input}  Design (D), Branch target ($B_i$)\\
\hspace*{\algorithmicindent} \textbf{Output} Test $T$ 
\end{flushleft}

\begin{algorithmic}[1]
\State $SS$ $\gets $SequenceIdentification($D$, $B_i$)
\State $<iD, TQ>$$ \gets$DesignInstrumentation($SS$, Design)
\State $T \gets$IncrementalConcolic($iD$,$TQ$)
\State Return $T$
\end{algorithmic}
\end{algorithm}

\vspace{-0.2in}
\subsection{Sequence Identification}

Algorithm~\ref{alg:sequence} shows the procedure for sequence detection for a given branch target. 
This algorithm consists of four major steps. The first step constructs the CFG for the design. This step can be performed using any existing Verilog language parser. Figure~\ref{fig:cfg} shows the CFG representation of the design in Listing 1. The next step extracts the branch condition for the target. This condition is an expression of the signals ($SE$). The third step uses {\it DependencySearch} function to recursively identify the assignment blocks that are relevant for each of the signal in $SE$. 
The final step returns the sequence of assignment blocks for activating the branch target. 

\begin{algorithm}[htp]
\footnotesize
\caption{Sequence Identification}\label{alg:sequence}
\begin{flushleft}

\hspace*{\algorithmicindent} \textbf{Input}  Design (D), Branch target ($B_i$)\\
\hspace*{\algorithmicindent} \textbf{Output} Sequence Stack ($SS$)
\end{flushleft}

\begin{algorithmic}[1]
\State $CFG$ $\gets $ConstructCFG(D)
\State $SE \gets$ GetSignalExpression($B_i$.condition)
\State $SS \gets $DependencySearch($CFG$, $SE$)
\State Return $SS$
\State

\Function{DependencySearch}{$CFG$, $SE$}
\For{each signal $A \in SE$}
\State $B_A$ $\gets $FindAssignmentBlock($CFG$, $A$)
\State $SS$.push($B_A$)
\State DependencySearch($CFG$, $A$)
\EndFor
\State \bf{Return} $SS$
\EndFunction
\end{algorithmic}
\end{algorithm}

\vspace{-0.1in}
\noindent {\bf Example 2:}
In Listing~\ref{lst:example1}, consider the target as line 37 where the block is ($B_{15}$) and this is represented in Figure~\ref{fig:cfg} as the ``Target". Line 1 of Algorithm~\ref{alg:sequence} produces three concurrent CFGs with inter-CFG edges in Figure~\ref{fig:cfg}. Line 2 of Algorithm~\ref{alg:sequence} produces the branch condition (line 36 in Listing~\ref{lst:example1}) as $SE \gets<r\_data$ == DATA$>$. This signal expression consists of one signal ($r\_data$) and one constant value ($DATA$). Since no action needed for $DATA$, the DependencySearch routine only tries to find the assignment block corresponding to signal $r\_data$. The signal $r\_data$ appears in one assignment (Line 25 in Listing~\ref{lst:example1}) where $r\_data$ is assigned the value of $mem[addr]$ in $CFG_2$ block $B_8$. The block $B_8$ is pushed into $SS$. Then the dependency search is executed for the signals $mem$ and $addr$. Since the $addr$ is a primary input, the search will not continue for $addr$.  An assignment exists for $mem$ in line 15 where $mem[addr]$ is assigned the value of $w\_data$ in $CFG_1$ block $B_3$. The block $B_3$ is pushed into $SS$. Since $w\_data$ is a primary input and there are no more assignments for $w\_data$, the recursion will end. Once the algorithm terminates, $SS$ will have $<B_3, B_8>$. The dependency search for $r\_data$ is shown in Figure~\ref{fig:cfg} using the two purple dotted lines. $\square$



\subsection{Design Instrumentation}

Algorithm~\ref{alg:branch} shows the procedure for branch generation for a given sequence set $SS$. As shown in the algorithm, breadth first search (BFS) is performed along the predecessors of the target block in the CFG (Intra-BFS) to extract the conditions to activate the target. Line 1 of the algorithm identifies the constraints for the target. For each sequence in the $SS$, it tries to identify the constraints using the similar intra-BFS (line 3). The constraints can have either resolved Boolean expressions or unresolved expressions. In the next step, constraints from the target are used to resolve the unresolved constraints of the sequence. First an intersection is performed between the unresolved constraints from the sequence and constraints from the target. The results of the intersection are the new resolved constraints for the sequence. If still some of the constraints are unresolved in the sequence, it  searches through dependencies to identify any dependent signals for the target. If any of the dependent signals are in the target constraints, the value of the target constraint is used to resolve the sequence constraint.



\noindent {\bf Example 3:} To identify the constraints for ``Target'' block ($B_{15}$ in Figure~\ref{fig:cfg} and line 37 in Listing 1), intra-BFS is performed in $CFG_3$. This search is represented using blue dotted lines in Figure~\ref{fig:cfg}. Intra-BFS for ``Target'' is $<B_{15}$, $B_{13}$, $B_{12}$, $B_{9}>$. Based on this traversal, we get the constraints to activate ``Target'' as $r\_en=1$, $w\_en=0$, $addr=ADDR$ and $r\_data=DATA$. Next, Intra-BFS is performed for the blocks in $SS$ ($<B_3, B_8>$). The constraints for $B_3$ are $r\_en=0$, $w\_en=1$, $mem=UR$, $addr=UR$ and $w\_data=UR$, and the constraints for $B_8$ are $r\_en=1$, $w\_en=0$, $mem=UR$, $addr=UR$ and $r\_data=UR$. Here, $UR$ means unresolved. There are three unresolved constrained for $B_3$. We can resolve the first constraint $addr=UR$ to $addr=ADDR$.   We need to search for dependencies to address the remaining two unresolved constraints ($mem$ and $w\_data$). The search of dependencies for $w\_data$ is shown in Figure~\ref{fig:cfg} using red dotted lines. $w\_data$ is assigned to $mem[addr]$ and $mem[addr]$ is assigned to $r\_data$. Once the search is complete, final dependency for $w\_data$ can be identified as $r\_data$. Since $r\_data$ is included the target constraints, $w\_data$ gets the value of $r\_data$. After discarding the unresolved constraints, the final values for $B_3$ are $r\_en=0$, $w\_en=1$, $addr=ADDR$ and $w\_data=DATA$, and for $B_8$ are $r\_en=1$, $w\_en=0$, $addr=ADDR$ and $r\_data=DATA$. $\square$


\begin{algorithm}[htp]
\footnotesize
\caption{Design Instrumentation}\label{alg:branch}
\begin{flushleft}

\hspace*{\algorithmicindent} \textbf{Input}  Design (D), CFG, Target ($B_i$), Sequence Stack ($SS$)\\
\hspace*{\algorithmicindent} \textbf{Output} Instrumented Design (iDesign), Target Queue ($TQ$)
\end{flushleft}

\begin{algorithmic}[1]
\State Target Constraints $TC \gets$ IntraBFS(CFG, $B_i$.block)
\For{each $S \in SS$ }
\State Sequence Constraints $SC \gets$IntraBFS(CFG, $S$)
\State $SC \gets $MODIFY($TC$, $SC$, CFG)
\State $TQ \gets $CreateBranch($SC$.resolved, D)
\State iDesign $\gets$ instrumentDesign(D, TQ)
\EndFor
\State {\bf Return} iDesign, $TQ$
\State

\Function{modify}{$TC$, $SC$, CFG}
\State $SC$.resolved $\gets SC$.unresolved $\cap~TC$
\For{each $cons \in SC$.unresolved}
    \State  Depend Signal $DS \gets $Search(CFG, $cons$.signal)
    \If{$DS \in TC$}
    \State $cons$.value $\gets TC$[$DS$].value
    \State $SC$.resolved $\gets SC$.resolved~$\cup~cons$
    \EndIf
\EndFor

\State {\bf Return} $SC$
\EndFunction
\end{algorithmic}
\end{algorithm}









In Algorithm~\ref{alg:branch}, for each of the sequences in $SS$,  conditional branches are created using the modified constraints (line 5) and these branches are embedded in the design. The newly created branches are stored in the $TQ$ (Target Queue) preserving the order in the $SS$. When the first sequence is removed from the $SS$, corresponding branch of that sequence is the first element to insert in the $TQ$. This process continues until $SS$ is empty. Finally, the modified design is instrumented (line 6). The goal of the instrumentation is to identify which path is executed by analyzing the simulation log. Specifically, we add print statements for all the branch conditions and end of the blocks by using a unique identifier.



\vspace{-0.1in}
\subsection{Incremental Concolic Testing}

\begin{figure}[t]
    \centering
    \includegraphics[width=0.8\columnwidth]{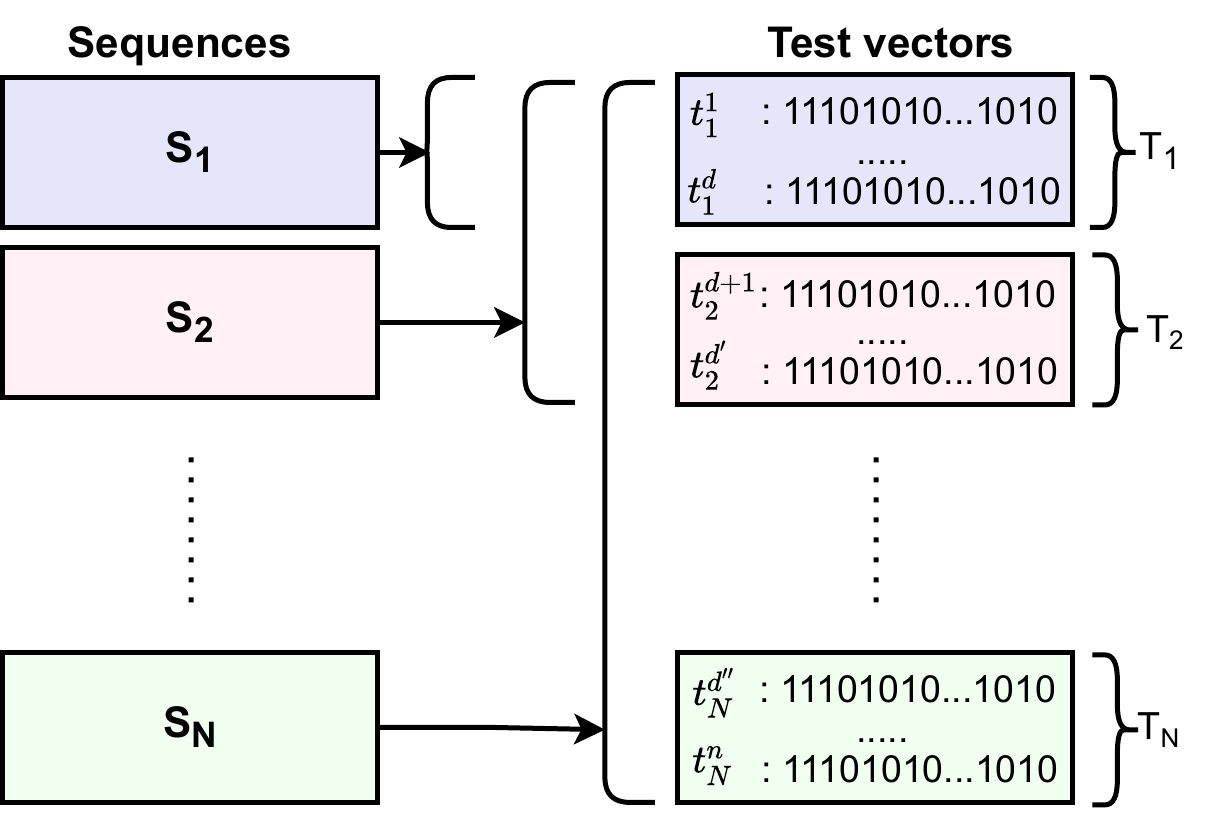}
    \vspace{-0.1in}
    \caption{Incremental test generation for a sequence set. $S_j$ is the j-th element of the sequence for the i-th branch $B_i^j$. }
    \label{fig:test}
        \vspace{-0.3in}
\end{figure}

In this section, we present the incremental concolic testing scheme to activate a set of sequence events in the preserved order. Figure~\ref{fig:test} presents a  pictorial representation of  incremental test generation. As shown in the figure, there are two sets: sequence set $<S_1$, $S_2$, ....., $S_N>$ and the corresponding test set $<T_1$, $T_2$, ...., $T_N>$. To activate a sequence $S_x$, the required test is $\sum_{k=1}^{x} T_k$. For example, $T_1$ can activate $S_1$ but to activate $S_2$, we need both $T_1$ and $T_2$. A test set is a combination of different test vectors. A test $T_x$ includes $\sum_{i=a}^{b} t^i_x$ where $a,b \leq n$ (unroll cycle). The test vectors in $T_1$ is $<t^1_1$, $t^2_1$, ...., $t^d_1>$ and the test vectors in $T_2$ is $<t^{d+1}_2$, $t^{d+2}_2$, ...., $t^{d'}_2>$.
\vspace{-0.1in}
\begin{lstlisting}[frame=single, caption={Branch creation for sequences},basicstyle=\footnotesize, label={lst:example2}, language=Verilog]
1.  if (r_en==1'b0 && w_en==1'b1 && 
     addr==ADDR  && w_data==DATA) begin
2.    $display("Target1") //B17
3.  end
4.  if (r_en==1'b1 && w_en==1'b0 && 
     addr==ADDR  && r_data==DATA) begin
5.    $display("Target2") //B19

\end{lstlisting}
\vspace{-0.05in}

Algorithm~\ref{alg:incremental} describes the incremental test generation using concolic testing to activate a sequence of events preserving the order of events. Specifically, the test generated to activate the current event is
used as the starting point to activate the next event in the
sequence. 
For each target in $TQ$, we run concolic testing while changing the test set and the starting clock cycle (line 4). For the first target, the test set ($T$) is generated randomly and it contains test vectors up to the unroll cycle ($n$). The first step of concolic framework is to calculate the distance from the target to all the blocks. From the target breadth-first traversal is performed in the direction along the predecessors. The distance is initialized to 0 and incremented by 1 when an edge traversal is completed. Next, path ($P$) is generated by simulating the design with test set $T$. All the alternate branches from the current path $P$ is selected as the next step. When selecting the alternate branches, the clock is set to a specific starting clock cycle value so that we only select the branches after the starting clock cycle value. The path up to the starting clock cycle is set and unchanged. Then the selected alternate branches are sorted using the distance and the clock value. This will lead to the most profitable alternate branch. Using the trace of $P$ and the chosen branch, constraint vector is generated. The constraint vector contains the value of the constraints for each of the clock cycles. Then the constraint vector is solved using a constraint solver. The constraint solver produces a new test set and this is used to simulate the design and get a new path. If the new path activates the target, the test set will be added to $T$. Also, the clock cycle of the selected branch will be set as the new starting clock cycle. Hence, the test set generated for the target will be preserved and used as a starting point to the next target in $TQ$.


\vspace{0.05in}
\noindent {\bf Example 4:} 
Target Queue ($TQ$) contains 2 branch targets $<B_{17}$, $B_{19}>$ which are shown in Listing~\ref{lst:example2}. Assume that the unroll cycle ($n$) is 10 and search $limit$ is 10 iterations. Concolic testing is used to activate the first branch ($B_{17}$) which is corresponding to writing a value to the memory. The $start$ value is 1 and a random test set is used as initial setting. If the tests to activate the target ($B_{17}$) is identified in unroll cycle 3, then the starting cycle is set as 4 for the next target ($B_{19}$). $\square$ 





\vspace{-0.1in}
\begin{algorithm}
\footnotesize
\caption{Incremental Concolic Testing}\label{alg:incremental}
\begin{flushleft}

\hspace*{\algorithmicindent} \textbf{Input} Design  (D), Target Queue ($TQ$),  Unrolled Cycles ($n$), $limit$\\
\hspace*{\algorithmicindent} \textbf{Output} Test Set $T = {T_1, T_2, . . . , T_N}$
\end{flushleft}

\begin{algorithmic}[1]
\State $T \gets $ Random Vectors
\State $start \gets $1
\For{each $target$ in $TQ$}
\State $T, start \gets $CONCOLIC(D, $target$, $T$, $start$)
\EndFor
\State \textbf{return} $T$
\State
\Function{Concolic}{Design, $target$, $T$, $start$}
\State Distance Set $DS \gets$ ComputeDistance($target$, Design)
\State Path $P$ $\gets$ Simulate($T$, Design)
\State $clock \gets start$ 
\While{$iteration$ $<$ $limit$} 
\State $AB$ $\gets$ AlternateBranch($P$, $DS$, $clock$)
\State $CV \gets $ BuildConstraints($AB$, $P$)
\State Test $t \gets$ SolveConstraints($CV$)
\color{black}

\State $P \gets$ Simulate($t$, Design)
\If{$P$ activates the $target$}
\State $T.add(t)$
\State $start \gets AB$.clock
\State Break
\EndIf
\EndWhile
\State \textbf{return} $T$, $start$
\EndFunction
\end{algorithmic}
\end{algorithm}
\vspace{-0.21in}

%% file: sections/experiment.tex
\section{Experiments}
\label{sec:experiments}

\subsection{Experimental Setup}

For  experimental evaluation, we have selected a re-configurable cache implementation,  IOb-Cache~\cite{roque2021iob}. It interfaces with a processor and main memory. For the case studies in Section~\ref{subsec:cases}, we have selected configurations as follows: (address width=16, data width=32, write back, LRU replacement, and 4-way set-associative design). With the above configuration, we flattened the IOb-Cache module eliminating its hierarchy with Yosys synthesis tool. The flattened RTL netlist is about 10,000 lines of code. 
This configuration is used for validation of different functional scenarios outlined in Section~\ref{subsec:cases}. In order to generate the abstract syntax tree of the RTL model, we use Icarus Verilog Target API~\cite{icarusverilog}. We use Yices SMT solver~\cite{dutertre2014yices} for solving constraints. Incremental concolic testing is implemented on top of the concolic testing framework~\cite{lyu2020scalable}. In order to ensure validity of the generated test vectors, we simulate the original design with the generated test and analyze the Value Change Dump (VCD) to confirm the activation of the target (corner case). We ran our experiments on Intel i7-5500U @3.0GHz CPU with 16GB RAM.


\begin{table*}
\centering
\footnotesize
\caption{Comparison of system-level target activation using \cite{Mukherjee:2015}, \cite{lyu2020scalable}, and our approach} 
\label{tab:verification}
\vspace{-0.1in}
\resizebox{2\columnwidth}{!}{%
\begin{tabular}{|c|c|c|c|c|c|c|c|c|r|r|} 
\hline
\multicolumn{1}{|c|}{\multirow{2}{*}{\textbf{Cases}}} & \multicolumn{1}{c|}{\multirow{2}{*}{\begin{tabular}[c]{@{}c@{}}\textbf{Unroll Cycles }\\\textbf{(Bound)}\end{tabular}}} & \multicolumn{3}{c|}{\textbf{EBMC~\cite{Mukherjee:2015}}} & \multicolumn{3}{c|}{\textbf{Concolic~\cite{lyu2020scalable}}} & \multicolumn{3}{c|}{\textbf{Our Approach}} \\ 
\cline{3-11}
\multicolumn{1}{|c|}{} & \multicolumn{1}{c|}{} & \multicolumn{1}{c|}{\textbf{Activated}} & \begin{tabular}[c]{@{}c@{}}\textbf{Memory}\\\textbf{~(MB)}\end{tabular} & \begin{tabular}[c]{@{}c@{}}\textbf{Time }\\\textbf{(s)}\end{tabular} & \multicolumn{1}{c|}{\textbf{Activated}} & \begin{tabular}[c]{@{}c@{}}\textbf{Memory}\\\textbf{~(MB)}\end{tabular} & \begin{tabular}[c]{@{}c@{}}\textbf{Time }\\\textbf{(s)}\end{tabular} & \multicolumn{1}{c|}{\textbf{Activated}} & \multicolumn{1}{c|}{\begin{tabular}[c]{@{}c@{}}\textbf{Memory }\\\textbf{(MB)}\end{tabular}} & \multicolumn{1}{c|}{\begin{tabular}[c]{@{}c@{}}\textbf{Time }\\\textbf{(s)}\end{tabular}} \\ 
\hline
1 & 20 & No & - & - & Yes & \multicolumn{1}{r|}{82.34} & \multicolumn{1}{r|}{20.13} & Yes & 20.00 & 14.55 \\ 
\hline
2 & 20 & No & - & - & No & - & - & Yes & 34.67 & 25.78 \\ 
\hline
3 & 50 & No & - & - & Yes & \multicolumn{1}{r|}{215.84} & \multicolumn{1}{r|}{50.67} & Yes & 67.89 & 20.78 \\ 
\hline
4 & 50 & No & - & - & No & - & - & Yes & 182.56 & 82.91 \\ 
\hline
5 & 20 & No & - & - & No & - & - & Yes & 19.78 & 14.43 \\ 
\hline
6 & 20 & No & - & - & No & - & - & Yes & 30.24 & 23.91 \\ 
\hline
7 & 20 & Yes & \multicolumn{1}{r|}{597.81} & \multicolumn{1}{r|}{2.01} & Yes & \multicolumn{1}{r|}{20.56} & \multicolumn{1}{r|}{4.93} & Yes & 15.23 & 4.81 \\ 
\hline
8 & 20 & No & - & - & No & - & - & Yes & 45.89 & 25.89 \\
\hline
\end{tabular}}
\vspace{-0.2in}
\end{table*}


\subsection{Corner Case Scenarios}\label{subsec:cases}

We have created different memory verification cases to validate corner case functionalities of the cache and memory. Specifically, we consider the following eight corner cases.

\vspace{0.05in}
\noindent {\bf Case 1:} Write a specific value to memory (Listing~\ref{lst:case1}). 
\vspace{-0.1in}
\begin{lstlisting}[frame=single, caption={Case 1}, basicstyle=\footnotesize, label={lst:case1}, language=Verilog]
1.if (ready == 1'b1)
2. if (wstrb == 1'b1)
3.  if(addr == 16'h1234)
4.   if(w_data == 32'hCAFEFEED) begin
5.    $display("Target")
6.  end
\end{lstlisting}

\noindent {\bf Case 2:} Read a specific data form a specific address (Listing~\ref{lst:case2}). This scenario is similar to the target in Listing 1.
    \vspace{-0.1in}
\begin{lstlisting}[frame=single, caption={Case 2}, basicstyle=\footnotesize, label={lst:case2}, language=Verilog]]
1.if (ready == 1'b1)
2. if (wstrb == 1'b0)
3.  if(addr == 16'h1234)
4.   if(r_data == 32'hCAFEFEED) begin
5.    $display("Target")
6.  end
\end{lstlisting}

\vspace{0.05in}
\noindent {\bf Case 3:} Back to back writes to the same address. We copied the entries in   Listing~\ref{lst:case1} for 5 times and changed the data values.

\vspace{0.05in}
\noindent {\bf Case 4:} Back to back reads from the same address. We copied the entries in   Listing~\ref{lst:case2} for 5 times and changed the data values. 
   
\vspace{0.05in} 
\noindent {\bf Case 5:} Write data to a boundary location in memory. We used the  Listing~\ref{lst:case1}, created two copies, and changed the address value to 4'h0000 and 4'hFFFF, respectively.
   
\vspace{0.05in} 
\noindent {\bf Case 6:} Read data from a boundary location in memory. We used the same Listing~\ref{lst:case2}, created two copies, and changed the address value to 4'h0000 and 4'hFFFF, respectively.
   
\vspace{0.05in} 
\noindent {\bf Case 7:} Verify front-end and back-end addresses for correct address translation as shown in Listing~\ref{lst:case7}. The specifc address translations are identified by analyzing the RTL models of front-end and back-end modules.
    
 
\vspace{-0.1in}   
\begin{lstlisting}[frame=single, caption={Case 7}, basicstyle=\footnotesize, label={lst:case7}, language=Verilog]]
1.if (addr == 16'h1234)
2. if (front_end.data_addr == addr[15:2])
3.  $display("Target1")
4.  end
5.if (addr == 16'h1234)
6. if (back_end.write_addr == addr[15:6])
7.  $display("Target2")
8.  end
\end{lstlisting}



\vspace{0.05in}
\noindent {\bf Case 8:} Verify cache hit for a specific memory read. As shown in Listing~\ref{lst:case9}, when the required write happens before the read, the cache hit should get triggered.
    
\vspace{-0.1in}
\begin{lstlisting}[frame=single, caption={Case 8}, basicstyle=\footnotesize, label={lst:case9}, language=Verilog]]
1.if (ready == 1'b1)
2. if (wstrb == 1'b0)
3.  if(addr==16'h1234 && r_data=32'hCAFEFEED)
4.   if(cache_memory.hit == 1'b1) begin
5.    $display("Target")
6.  end
\end{lstlisting}



\subsection{Results}
\label{subsec:results}
In this section, we present the results of our case study. We compare our approach with EBMC~\cite{Mukherjee:2015} and the concolic testing framework presented in~\cite{lyu2020scalable}. EBMC is a state-of-the art formal verification framework that uses bounded model checking. The concolic testing  framework~\cite{lyu2020scalable} is state-of-the-art in activating RTL branch statements using concolic testing. The number of unrolled cycles are determined based on the complexity of the scenarios. This can be achieved by starting from a reasonable number of unroll cycles and increment until the scenarios are covered. The number of unroll cycles is analogous to the bound determination for bounded model checking. We set the bound for EBMC to be equal to the number of unroll cycles for concolic testing.

The corner case activation results at system level are shown in Table~\ref{tab:verification}. The first column represents different corner case scenarios outlined in Section~\ref{subsec:cases}. The second column provides the unroll cycles (bound for EBMC). For each approach, we provide information about if the target (corner case) is activated (Yes or No) within the bound, and if yes, what is the memory requirement (in MB) and run time (in seconds). As shown in Table~\ref{tab:verification}, EBMC only covers one scenario, and concolic~\cite{lyu2020scalable} covers only three scenarios. Our approach successfully covered all the 8 scenarios. EBMC is expected to fail for most of the scenarios due to state space exploitation problem. The concolic framework in~\cite{lyu2020scalable} activates some of the branches, however, when dealing with  contradictory and complex sequences, it fails to activate the target due to path explosion problem (\cite{lyu2020scalable} selects branches based on the distance heuristics).


\begin{table}[htp]
\centering
\footnotesize
\vspace{-0.1in}
\small
\caption{Memory(MB) and time(s) taken to verify Case 2.} 
\label{tab:result1}
\vspace{-0.1in}
\resizebox{\columnwidth}{!}{%
\begin{tabular}{|c|ccc|ccc|}
\hline
\multirow{2}{*}{\textbf{\begin{tabular}[c]{@{}c@{}}Unroll\\ cycles\end{tabular}}} & \multicolumn{3}{c|}{\textbf{Concolic~\cite{lyu2020scalable}}} & \multicolumn{3}{c|}{\textbf{Our Approach}} \\ \cline{2-7} 
 & \multicolumn{1}{c|}{\textbf{Activated}} & \multicolumn{1}{c|}{\textbf{Mem}} & \textbf{Time} & \multicolumn{1}{c|}{\textbf{Activated}} & \multicolumn{1}{c|}{\textbf{Mem}} & \textbf{Time} \\ \hline
10 & \multicolumn{1}{c|}{No} & \multicolumn{1}{r|}{52.4} & 29.92 & \multicolumn{1}{c|}{Yes} & \multicolumn{1}{r|}{10.9} & 0.59 \\ \hline
20 & \multicolumn{1}{c|}{No} & \multicolumn{1}{r|}{86.3} & 70.59 & \multicolumn{1}{c|}{Yes} & \multicolumn{1}{r|}{11.4} & 1.75 \\ \hline
30 & \multicolumn{1}{c|}{No} & \multicolumn{1}{r|}{121.2} & 137.25 & \multicolumn{1}{c|}{Yes} & \multicolumn{1}{r|}{12.9} & 7.09 \\ \hline
40 & \multicolumn{1}{c|}{No} & \multicolumn{1}{r|}{154.8} & 225.37 & \multicolumn{1}{c|}{Yes} & \multicolumn{1}{r|}{12.1} & 6.22 \\ \hline
50 & \multicolumn{1}{c|}{Yes} & \multicolumn{1}{r|}{164.6} & 286.09 & \multicolumn{1}{c|}{Yes} & \multicolumn{1}{r|}{13.1} & 11.75 \\ \hline
\end{tabular}}
\vspace{-0.15in}
\end{table}

To understand the limitation of the state-of-the-art RTL concolic framework in~\cite{lyu2020scalable}, we only consider the iob\_ram module with `Case 2' and compare the memory and time requirements. The experimental results are shown in Table~\ref{tab:result1}. The concolic framework in~\cite{lyu2020scalable} was able to activate the target (Case 2) only when unrolled for 50 cycles whereas our approach is able to activate the branch in 10 unroll cycles. The performance improvement of our approach compared to~\cite{lyu2020scalable} in terms of time and memory is 24 times and 12 times, respectively. It also highlights that state-of-the-art can activate corner cases if the design is sufficiently unrolled, which can be infeasible for industrial designs since various components in concolic testing (e.g., constraint solver) may not be able to handle such a large number of constraints. 

%% file: sections/conclusion.tex
\vspace{-0.1in}
\section{Conclusion}
\label{sec:conclusion}
Concolic testing provides a scalable test generation framework using an effective combination of simulation and formal methods. While it is promising for branch coverage in register-transfer level (RTL) deigns, it cannot activate complex corner cases such as hard-to-activate branches. 
We have developed an incremental concolic testing framework to cover such corner case scenarios in RTL models. Specifically, this paper made three important contributions. First, we show that a complex branch condition can be decomposed as a sequence of easy-to-activate events by traversing respective control and data flow graphs. Next, we map the branch coverage problem to the coverage of a sequence of events such that the test generated to activate the current event can be used as the starting point for activating the next event in the sequence. Finally, we have developed an efficient algorithm to cover the sequence of events by iterative invocation of concolic testing. Our experimental results demonstrated that our approach can be used to generate directed tests to cover complex branch targets in modern memory designs, while state-of-the-art methods fail to activate them.